\documentclass{article}

\usepackage[final]{nips_2018}

\usepackage[utf8]{inputenc} 
\usepackage[T1]{fontenc}    
\usepackage{hyperref}       
\usepackage{url}            
\usepackage{booktabs}       
\usepackage{amsfonts}       
\usepackage{nicefrac}       
\usepackage{microtype}      
\usepackage{graphicx}
\setcitestyle{comma,square,numbers}

\title{Real-Time Sleep Staging using Deep Learning on a Smartphone for a Wearable EEG}

\author{Abhay Koushik
\\
  MIT Media Lab\\
  Cambridge, MA USA \\
  \texttt{koushika@mit.edu}
  \And
  Judith Amores\\
  MIT Media Lab\\
  Cambridge, MA USA\\
  \texttt{amores@mit.edu}
  \And
  Pattie Maes\\
  MIT Media Lab\\
  Cambridge, MA USA\\
  \texttt{pattie@media.mit.edu}
}

\begin{document}

\maketitle

\begin{abstract}
We present the first real-time sleep staging system that uses deep learning without the need for servers in a smartphone application for a wearable EEG. We employ real-time adaptation of a single channel Electroencephalography (EEG) to infer from a Time-Distributed 1-D Deep Convolutional Neural Network. Polysomnography (PSG)\textemdash the gold standard for sleep staging, requires a human scorer and is both complex and resource-intensive. Our work demonstrates an end-to-end on-smart phone pipeline that can infer sleep stages in just single 30-second epochs, with an overall accuracy of 83.5\% on 20-fold cross validation for five-class classification of sleep stages using the open Sleep-EDF dataset.  
\end{abstract}

\section{Introduction and Background}

Having a proper night of sleep and regular circadian rhythm is crucial for physical, mental and social well-being \cite{simon2018sleep,wulff2010sleep}. Sleep facilitates learning, memory-consolidation and emotion processing\cite{asfestani2018overnight,kurdziel2018sleep}. Identification of sleep stages is important not only in diagnosing and treating sleep disorders but also for understanding the neuroscience of healthy sleep. Polysomnography (PSG) is used in hospitals to study sleep and diagnose sleep disorders. It is considered the gold standard for sleep staging and involves recording of multiple electrophysiological signals from the body such as brain activity using EEG, heart rhythm through Electrocardiography (ECG), muscle tone through Electromyography (EMG) and eye-movement through Electrooculography (EOG). PSG is a tedious procedure which requires skilled sleep technologists in a laboratory setting. Since EEG is the most reliable signal for sleep staging, automation of PSG can be achieved through accurate classification of EEG\cite{lucey2016comparison}. Research in Deep Learning\cite{lecun2015deep,krizhevsky2012imagenet} has led to many efficient algorithms to classify different kinds of data including bio-medical and physiological signals. In this paper, we focus on developing a real-time sleep staging application using Time-distributed 1-D Deep Convolutional Neural Networks to classify the five sleep stages in a comfortable environment. As per the new AASM rules\cite{berry2012aasm}, these stages are\textemdash Wake, Rapid-Eye-Movement (REM) and Non-Rapid-Eye-Movement (N-REM) stages N1, N2, and N3. We make use of a single channel EEG recorded through a modified research version of the Muse headband\cite{krigolson2017choosing}. The headband is flexible and can be comfortably used while sleeping. It has 5 electrodes, namely, AF7, AF8, TP9, TP10 and reference Fpz, for brain activity measurements.
\paragraph{Importance of mobile systems for sleep staging}
 PSG requires a minimum of 22 wires attached to the body in order to monitor sleep activity. The complexity of this setup requires sleeping in a hospital or laboratory with an expert monitoring and scoring signals in real-time. This results in an unnatural and disturbed night of sleep for the subject which may not only affect the diagnosis but also causes sub-optimal utilization of time and energy resources for recording and scoring and is as such highly undesirable. There is significant development in research on automating sleep staging with wireless signals\cite{zhao2017learning} and more compact, wearable devices\cite{casson2010wearable,sano20170182}. Nevertheless, none of these systems implements a five-stage classification of sleep in real-time.
 
The goal of our research was to simplify and reliably automate PSG on-smart phone in just unit 30-second non-overlapping epochs for automatic real-time interventions during experiments on sleep stages and cognition as the minimum time-resolution of a single sleep score by an expert is 30 seconds. Automated classification is achieved through adaptation of a Time-Distributed Deep Convolutional Neural Network model. Simplification is achieved by developing TensorFlow Lite Android application that uses only a single channel recording from a wearable EEG. We have also designed a friendly user interface that visualizes sleep stages and raw EEG data with real-time statistics about accuracy. Our app connects via Bluetooth Low Energy (BLE) to the flexible EEG headband thus making it portable and not restricted to laboratory and hospital use.

\section{Related work}
Automatic analysis and sleep scoring using multi-layer Neural  Networks \cite{schaltenbrand1996sleep} was done as early as 1996 using 3 channels of physiological data, namely EEG, EOG and EMG. This involved power spectral density calculations for feature extraction from raw EEG which required a tedious laboratory setting to collect reliable data through these channels. More recent work has looked into creating portable sleep scoring systems, such as the work by Zhang et al. \cite{zhang2012rass}, that uses pulse, blood oxygen and motion sensors to predict sleep stages. In their paper, they do not detect sleep stages N1 and N2 separately, and N1 is usually the hardest one to predict. The authors already mention that these results cannot provide equally high accuracy as compared to the EEG and EOG signals of PSG. The same limitations apply to the work by Zhao et al\cite{zhao2017learning}. Our work achieves reliable accuracy by using only one channel from a wearable EEG, and overcomes the complexity of recording multiple signals. 

Our model is based on Time-Distributed Deep Convolutional Neural Networks\cite{cvxtz2018eeg} and is inspired by the DeepSleepNet from Supratak et al.\cite{supratak2017deepsleepnet}. DeepSleepNet makes use of representation learning with a Convolutional Neural Network (CNN) followed by sequence residual learning using Bidirectional-Long Short Term Memory cells (Bi-LSTM). The major drawback of this network is that it requires 25 epochs of raw EEG data to be fed in together to obtain 25 labels. This is mainly because of the Bi-LSTM which relies on large temporal sequences to achieve better accuracy.

State-of-the art network model\textemdash SeqSleepNet\cite{phan2018seqsleepnet} processes multiple epochs and outputs the sleep labels all at once using end-to-end Hierarchical Recurrent Neural Networks. This uses all 3 channels\textemdash namely, EEG, EMG and EOG in order to give the best overall accuracy of 87.1\% on the MASS dataset\cite{o2014montreal}. CNN models by Sors et al.\cite{sors2018convolutional} and Tsinalis et al.\cite{tsinalis2016automatic}, as well as SeqSleepNet and DeepSleepNet all use longer temporal sequences for inference\textemdash 4, 5, 10 and 25 raw EEG epochs of 30 seconds respectively. We overcome this limitation by using Time-Distributed Deep CNN to predict single 30-second epochs with real-time adaptation from wearable EEG.

Flexibility of this wearable also makes it more preferable than the bulky system used by Lucey et al\cite{lucey2016comparison}. The smart-phone based nature of our sleep-staging application overcomes the need for a client-server architecture as used in Dreem headband\cite{patanaik2018end}. Our TensorFlow-Lite mobile application can also be adapted to other types of EEG devices for real-time settings.
\section{Methods and Materials}

\subsection{Dataset description and pre-processing}
We used the expanded Sleep-EDF database from Physionet-bank\cite{kemp2000analysis}. Single channel EEG (Fpz-Cz at 100Hz) of 20 subjects are divided into training set of 33 nights and validation set of 4 nights. Together, they contain non-overlapping nights of 19 subjects for 20 fold-cross validation. The non-overlapping test set contains 2 nights (1 subject).  We remove the extra wake states before and after half an hour of sleep as described in the DeepSleepNet\cite{supratak2017deepsleepnet}. We excluded \textit{MOVEMENT} and \textit{UNKNOWN} stages, and combined N4 and N3 to follow the five-stage classification as per the new AASM rules\cite{berry2012aasm}.

\subsection{Model architecture and training}
\begin{figure}[ht]
    \setlength{\fboxsep}{0pt}
    \setlength{\fboxrule}{1pt}
    \centering
    \includegraphics[width=12cm,height=0.77cm]{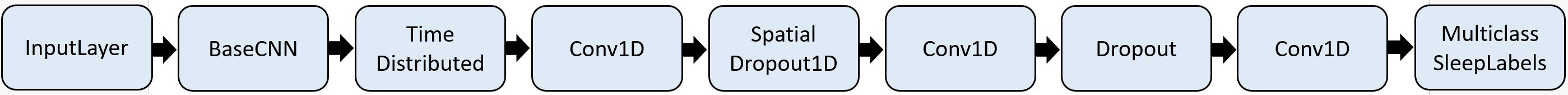}
    \caption{Model architecture used for training and inference}
    \label{model_cnn}
\end{figure}
Our model architecture is described in Figure \ref{model_cnn}. The Base-CNN has 3 repeated sets of two 1-D convolutional (Conv1D) layers, 1-D max-pooling and spatial dropout layers. This is followed by two Conv1D, 1-D global max-pooling, dropout and dense layers. We finally have a dropout layer as the output of Base-CNN. 30-second epochs of normalized raw EEG at 100Hz is fed into the Time-Distributed Base-CNN model\cite{cvxtz2018eeg} as described in the Figure \ref{model_cnn}. All Conv1D layers use Rectified-Linear-Units (ReLU) activation. The training uses an Adam optimizer of 0.001 with an initial learning rate of $e^{-3}$ which is reduced each time the validation accuracy plateaus using ReduceLROnPlateau Keras Callbacks.
\subsection{EEG adaptation and experiments} \paragraph{Pre-processing data from wearable EEG} Real-time brain activity from the flexible EEG headband is streamed via BLE to the smartphone application. Raw EEG from Af7 channel is down-sampled to 100Hz at the end of each 30-second epoch before feeding into the network. 

\begin{figure}[h!]
    \setlength{\fboxsep}{0pt}
    \setlength{\fboxrule}{1pt}
    \centering
    \includegraphics[width=13cm]{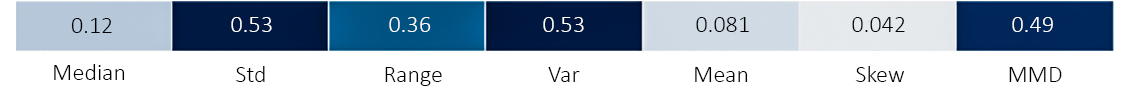}
    \caption{The heat-map shows the mutual importance of different statistical features of raw EEG. We choose the default of 3 nearest neighbours as the parameter for the mutual\_info\_classif function for automated feature-selection in scikit-learn machine learning library}
    \label{sleep_features}
\end{figure}

\paragraph{EEG real-time adaptation} 
Since the raw EEG recording instrument used for training is different from the testing instrument, we adapt EEG using a Z-score scaling with wake-stage standard deviation for calibration. The main reason for choosing standard deviation of EEG over any other metrics is highlighted in Figure \ref{sleep_features}. We estimate the mutual information for a discrete sleep stage variable. We use the set of statistical features shown in the diagram as input calculated every 30 seconds and the corresponding sleep-stage labels as output of the dataset. Final feature shown is Min-Max-Distance (MMD) as described by Aboalayon et al.\cite{aboalayon2016sleep}, calculated as a sum over euclidean distances between minimum and maximum EEG values with 1 second sliding window. 

We calculated the statistical feature importance of raw EEG for 3 randomly selected nights. The standard deviation clearly has the major role with average relative importance of 53.33 percent over other features for the classification of sleep stages. Z-score calibration of wake stage employs this important feature, hence, using this method for 30-second epoch adaptation makes the raw-EEG both instrument-independent and subject-independent as long as signal is noise-reduced.

\section{Results} 

\begin{figure}[ht]
    \centering
    \includegraphics[height=6cm,width=9cm]{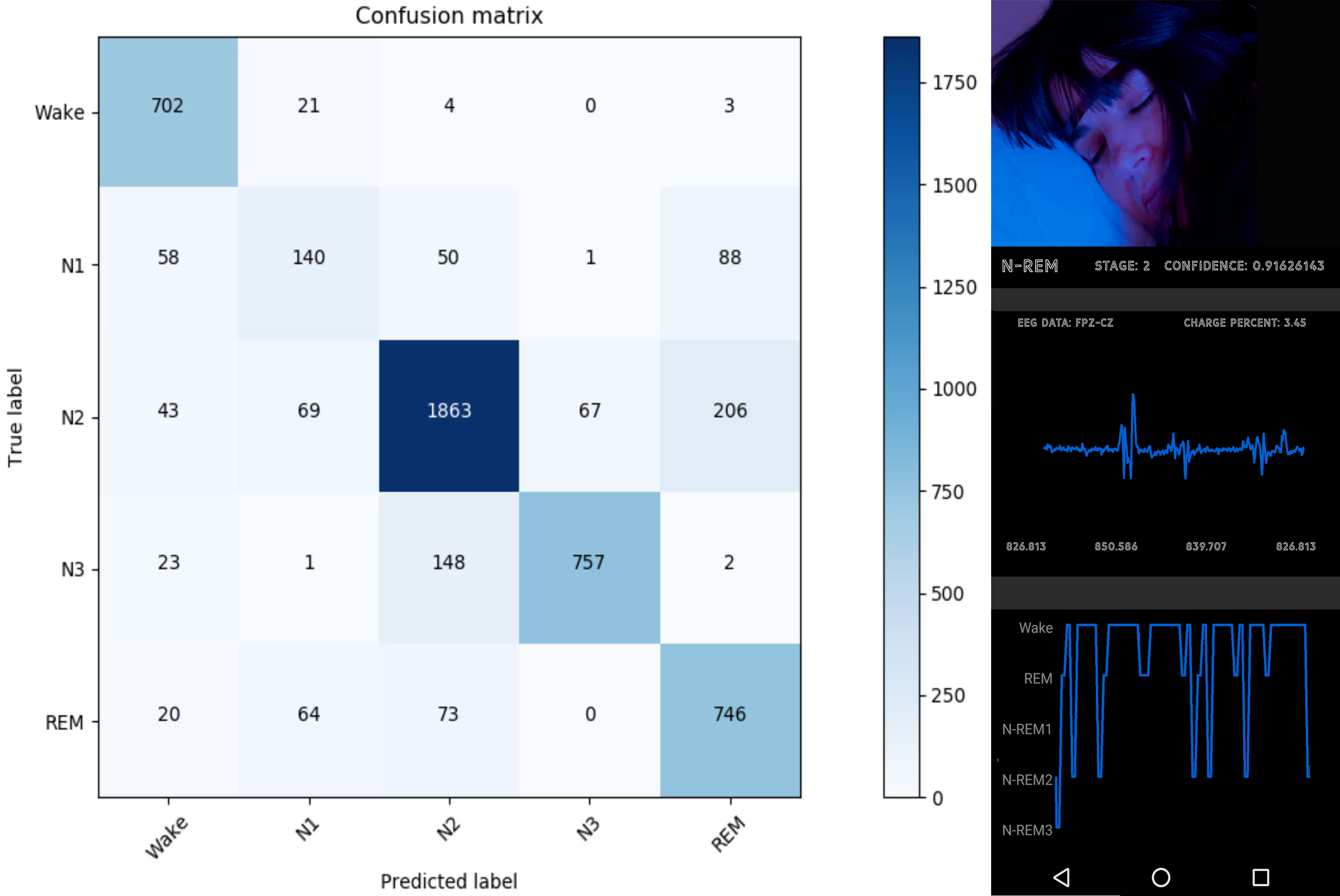}
    \caption{Confusion matrix (left). Snapshot of the headband and android application (right)}
    \label{sleep-multiple}
\end{figure}

\begin{figure}[h!]
    \centering
    \includegraphics[height=5cm,width=12cm]{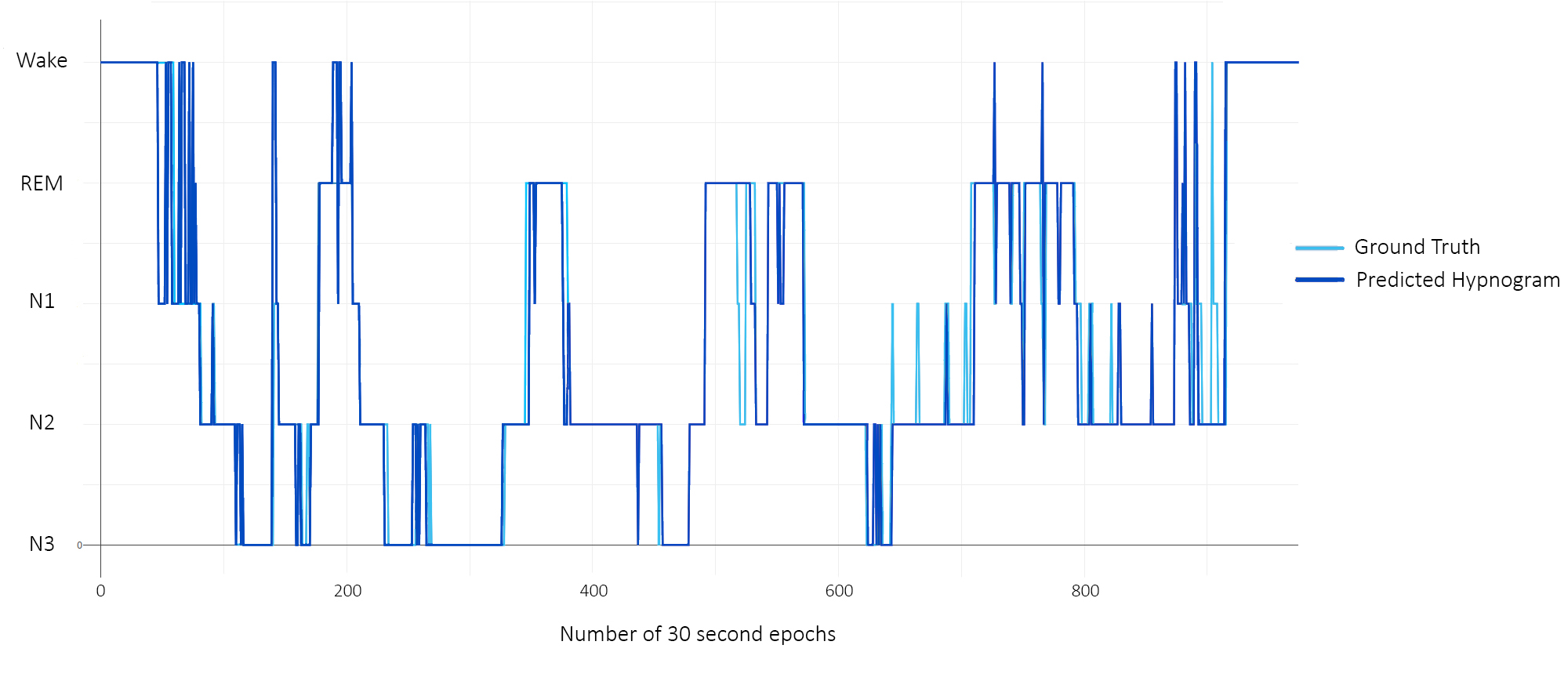}
    \caption{Comparison of the ground truth and predicted hypnograms of one full night}
    \label{hypnogram}
\end{figure}

Our model has an overall accuracy of 83.5\% for 20-fold cross validation of five-stage classification. The accuracy for nights from the test data ranges from 72\%(worst-case). This model achieves reliable accuracy given that the overall IRR (Inter-Rater-Reliability)\cite{danker2009interrater} among human experts scoring sleep recordings reported was about 80\% (Cohen’s $\kappa$ = 0.68 to 0.76). 
\begin{table}[ht]
  \caption{Five-class classification report for sleep staging on 5 test nights}
  \label{class-table}
  \centering
  \begin{tabular}{llllll}
    \toprule
    Label    & Sleep Stage     & Precision     & Recall     & F1-score     & Support \\
    \midrule
    0    & Wake     & 0.83    & 0.96    & 0.89    & 730 \\
    1    & N1       & 0.47    & 0.42    & 0.44    & 337 \\
    2    & N2       & 0.87    & 0.83    & 0.85    & 2248 \\
    3    & N3       & 0.92    & 0.81    & 0.86    & 931 \\
    4    & REM      & 0.71    & 0.83    & 0.77    & 903 \\
    \midrule
    Micro    & average    & 0.82    & 0.82    & 0.82    & 5149 \\
    Macro    & average    & 0.76    & 0.77    & 0.76    & 5149 \\
    Weighted    & average    & 0.82    & 0.82    & 0.82    & 5149 \\
    \bottomrule
  \end{tabular}
\end{table}
Table\ref{class-table} describes the precision, recall, F1-score and support of all the five sleep stages on predictions from 5 test nights. The corresponding accuracy of 81.72\% and F1-score of 76.23\% was obtained. In addition, micro, macro and weighted average of these metrics are also calculated in order to give a better statistical understanding. The confusion matrix for the same night is shown in the left part of Figure \ref{sleep-multiple}. N1 stage shows the poorest agreement because of the absence of an occipital electrode\cite{lucey2016comparison}.
The Figure \ref{hypnogram} shows the difference between the hypnogram of one full-night predicted by the model and the ground truth hypnogram as labeled in the dataset. The mobile application which deploys this deep learning model is built using TensorFlow Lite. The raw data from the wearable EEG is wirelessly streamed to the phone and updated every 30 seconds with its correspondent sleep stage and confidence value. We successfully validated real-time Rapid-Eye-Movement (REM) detection using our wearable headband by re-creating the same closed, lateral eye movement during wakefulness. We have also simulated and successfully validated jaw clenching and blinks during wakefulness.
\section{Conclusion and future scope}
This work demonstrates an end-to-end mobile pipeline for the fastest real-time sleep-staging by adaptation of a wearable EEG. With the development of the new mobile TensorFlow Lite application, we achieve automated sleep staging without the need of servers, in a portable way that can be used anywhere, including the home. The application is versatile as it can be adapted to take in single channel(Fpz-Cz) recordings from any wearable EEGs. We aim to use this work for real-time interventions using Brain Computer Interfaces (BCI) for applications in Human Computer Interaction (HCI), such as wearable olfactory interfaces \cite{amores2018bioessence,amores2017essence}, real-time audio-neural feedback\cite{schutze2015difficulty,dafna2018sleep} and sleep-based enhancement of learning and memory\cite{rasch2007odor,andrillon2017formation}.
\newpage
\bibliographystyle{plain}
\small{

}
\newpage
\section{Appendix}

\begin{figure}[ht]
    \setlength{\fboxsep}{0pt}
    \setlength{\fboxrule}{1pt}
    \centering
    \includegraphics[width=8cm]{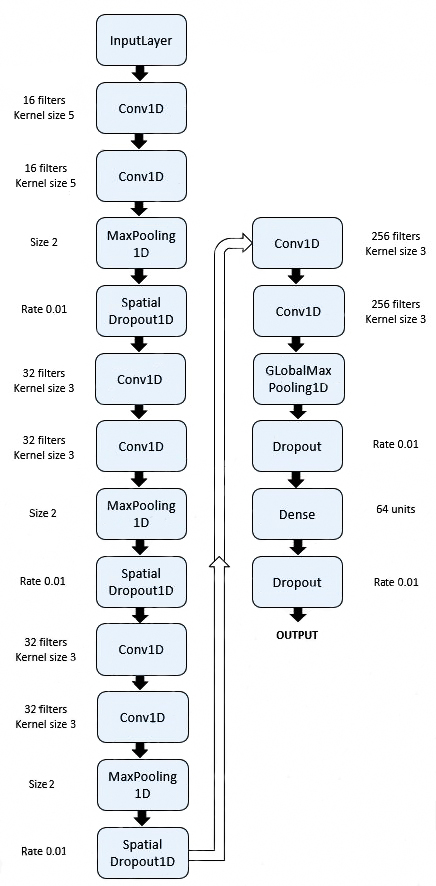}
    \caption{Architecture of the Base-CNN model}
    \label{base_cnn}
\end{figure}

The Base-CNN model used in our work is described by the Figure \ref{base_cnn}. All dropout layers have a rate of 0.01, pooling layers have a size of 2 and the model is compiled with an Adam optimizer of 0.001. The corresponding parameters of each of the layers are given alongside for reference.

\end{document}